\documentclass{mn2e}

\newcommand{\Msolar}{\mbox{\,$\rm M_{\odot}$}}        
\newcommand{\Lsolar}{\mbox{\,$\rm L_{\odot}$}}        

\newcommand{\kms}{km~s$^{-1}$}
\usepackage{graphicx}
\usepackage{color}
\usepackage{lscape}

\begin{document}
\title[]{Stellar and gaseous velocity dispersions in type II AGNs
at $0.3<z<0.83$ from the Sloan Digital Sky Survey}

\author[W. Bian, Q. Gu, Y. Zhao, L. Chao, Q. Cui]
{W.Bian$^{1,2}$, Q. Gu$^{3}$\thanks{Visiting Scholar,
Harvard-Smithsonian Center for Astrophysics, 60 Garden St.,
Cambridge, MA 02138, USA}, Y. Zhao$^{3}$, L. Chao$^{1}$ and Q. Cui$^{1}$ \\
$^{1}$Department of Physics and Institute of Theoretical Physics,
Nanjing Normal University, Nanjing
210097, China\\
$^{2}$Key Laboratory for Particle Astrophysics, Institute of High
Energy Physics, Chinese Academy of Sciences, Beijing 100039,
China\\
$^{3}$Department of Astronomy, Nanjing University, Nanjing
210093, China\\
}

\maketitle

\begin{abstract}
We apply the stellar population synthesis code by Cid Fernandes et
al. to model the stellar contribution for a sample of 209 type II
AGNs at $0.3<z<0.83$ from the Sloan Digital Sky Survey. The
reliable stellar velocity dispersions ($\sigma_{*}$) are obtained
for 33 type II AGNs with significant stellar absorption features.
According to the $L_{\rm [O III]}$ criterion of $3\times 10^{8}
\Lsolar$, 20 of which can be classified as type II quasars. We use
the formula of Greene \& Ho to obtain the corrected stellar
velocity dispersions ($\sigma_{*}^c$). We also calculate the
supermassive black holes masses from $\sigma_{*}^c$ in these
high-redshift type II AGNs. The [O III] luminosity is correlated
with the black hole mass (although no correlation between the
extinction-corrected [O III] luminosity and the black hole mass),
and no correlation is found between the Eddington ratio and the [O
III] luminosity or the corrected [O III] luminosity. Three sets of
two-component profile are used to fit multiple emission
transitions ([O III]$\lambda\lambda$4959, 5007 and [O
II]$\lambda\lambda$3727, 3729) in these 33 stellar-light
subtracted spectra. We also measure the gas velocity dispersion
($\sigma_{g}$) from these multiple transitions, and find that the
relation between $\sigma_{g}$ and $\sigma_{*}^c$ becomes much
weaker at higher redshifts than in smaller redshifts. The
distribution of $<\sigma_{g}/\sigma_{*}^c>$ is $1.24\pm0.76$ for
the core [O III] line and $1.20\pm0.96$ for the [O II] line, which
suggests that $\sigma_{g}$ of the core [O III] and [O II] lines
can trace $\sigma_{*}^c$ within about 0.1 dex in the logarithm of
$\sigma_{*}^c$. For the secondary driver, we find that the
deviation of $\sigma_{g}$ from $\sigma_{*}^c$ is correlated with
the Eddington ratio.
\end{abstract}

\section{INTRODUCTION}
Recent advances in the study of normal galaxies and active
galactic nuclei (AGNs) are ample observational evidence for the
existence of central supermassive black holes (SMBHs) and the
relationship between SMBHs and bulge properties of host galaxies
(Gebhardt et al. 2000; Ferrarese \& Merritt 2000; Tremaine et al.
2002; Begelman 2003; Shen et al. 2005). We can use the stellar
and/or gaseous dynamics to derive the SMBHs masses in nearby
inactive galaxies. However, it is  much more difficult  for the
case of AGNs. With the broad emission lines from broad-line
regions (BLRs)(e.g. H$\beta$, Mg II, CIV; H$\alpha$), the
reverberation mapping method and the empirical size-luminosity
relation can be used to derive the virial SMBHs masses in AGNs
(Kaspi et al. 2000; Vestergaard 2002; McLure \& Jarvis 2002; Wu et
al. 2004; Greene \& Ho 2006a).  It has been found that nearby
galaxies and AGNs follow the same tight correlation between the
central SMBHs masses ($M_{\rm BH}$) and stellar bulge velocity
dispersion ($\sigma_{*}$) (the $M_{\rm BH}-\sigma_{*}$ relation)
(Nelson et al. 2001; Tremaine et al. 2002; Greene \& Ho 2006a,
2006b), which also implied that the mass from reverberation
mapping method is reliable.

According to AGNs unification model (e.g. Antonucci 1993; Urry \&
Padovani 1995), AGNs can be classified into two classes depending
on whether the central engine and BLRs are viewed directly (type I
AGNs) or are obscured by circumnuclear medium (type II AGNs). In
type I AGNs, by using the broad emission lines from BLRs  (the
reverberation mapping method or the empirical size-luminosity
relation), we can derive virial SMBHs masses. It is not easy to
study their host galaxies because their optical spectra are
dominated by the non-stellar emission  from the central AGNs
activity. This is especially true for luminous AGNs, where the
continuum radiation from central source outshines the stellar
light from the host galaxy.

In type II AGNs, the obscuration of BLRs makes both the
reverberation mapping method and the empirical size-luminosity
relation infeasible to derive SMBHs masses. However, we can use
the well-known $M_{\rm BH}-\sigma_{*}$ relation to derive SMBHs
masses if we can accurately measure the stellar bulge velocity
dispersion ($\sigma_{*}$). There are mainly two different
techniques to measure $\sigma_{*}$, one is the "Fourier-fitting"
method (Sargent et al. 1977; Tonry \& Davis 1979), the other is
the "Direct-fitting" method (Rix \& White 1992; Greene \& Ho 2006b
and reference therein). These years it has been successful to
derive $\sigma_{*}$ through fitting the observed stellar
absorption features, such as Ca H+K $\lambda \lambda$ 3969, 3934,
Mg Ib$\lambda \lambda$ 5167, 5173, 5184 triplet, and Ca II$\lambda
\lambda$ 8498, 8542, 8662 triplet, etc,  with the combination of
different stellar template spectra broadened by a Gaussian kernel
(e.g. Kauffmann et al. 2003; Cid Fernandes et al. 2004a; Greene \&
Ho 2006b).

On the other hand, Nelson \& Whittle (1996) find that the gaseous
velocity dispersion ($\sigma_{g}$) of [O III]$\lambda$5007 from
the narrow-line regions (NLRs) is nearly the same as $\sigma_{*}$
for a sample of 66 Seyfert galaxies, and suggest that the gaseous
kinematics of NLRs be primarily governed by the bulge
gravitational potential. Nelson (2001) find a relation between
$M_{\rm BH}$ and $\sigma_{\rm [O III]}$ (the [O III]$\lambda$5007
velocity dispersion) for AGNs, very similar to the relation of
$M_{\rm BH}-\sigma_{*}$, although with more scatter, which
strongly suggests that $\sigma_{g}$ can be used as a proxy for
$\sigma_{*}$. For lower-redshift type II AGNs with $0.02 < z <
0.3$, Kauffmann et al. (2003) have investigated the properties of
their hosts from the Sloan Digital Sky Survey (SDSS) Data Release
One (DR1), measured $\sigma_{*}$ and estimated the SMBHs masses
from $\sigma_{*}$ (Brinchmann et al. 2004). By using this sample,
Greene \& Ho (2005) have measured the gaseous velocity dispersion
($\sigma_{\rm g}$) from multiple transitions ([O II]
$\lambda$3727, [O III] $\lambda$5007, and [S II]
$\lambda\lambda$6716, 6731) and compared $\sigma_{*}$ and
$\sigma_{g}$. They find that $\sigma_{g}$ from these multiple
transitions trace $\sigma_{*}$ very well, although some emission
features they show considerable scatters.

Type II quasars are the luminous analogs of low-luminosity type II
AGNs (such as Seyfert 2 galaxies). The obscuration of BLRs makes
quasars appear to be type II quasars (obscured quasars). Some
methods have been used to discover type II quasars, but only a
handful have been found.  Recently, Zakamsa et al. (2003) present
a sample of 291 type II AGNs at redshifts $0.3 < Z < 0.83$ from
the SDSS spectroscopic data. About half are type II quasars if we
use the [O III] $\lambda$5007 line luminosity to  represent the
strength of the nuclear activity. What is the $\sigma_{*} -
\sigma_{g}$ relation for type II quasars? And what are their SMBHs
masses and the Eddington ratios $L_{\rm bol}/L_{\rm Edd}$ (i.e.
the bolometric luminosity as a fraction of the Eddington
luminosity)?

Here we use the sample of Zakamsa et al. (2003) to study these
questions for type II quasars. In section 2, we introduce the data
and the analysis. Our results and discussion are given in Sec. 3.
All of the cosmological calculations in this paper assume
$H_{0}=70 \rm {~km ~s^ {-1}~Mpc^{-1}}$, $\Omega_{M}=0.3$,
$\Omega_{\Lambda} = 0.7$.

\section{data and analysis}
With SDSS, Zakamsa et al. (2003) presented a sample of 291 type II
AGNs at redshifts $0.3 < z < 0.83$. We downloaded these spectra
from SDSS Data Release Four (DR4) and the spectra for 202 type II
AGNs at redshifts $0.3 < z < 0.83$ are obtained. SDSS spectra
cover 3800-9200 $\AA$, with a resolution ($\lambda/\Delta
\lambda$) of $1800 < R < 2100$ and sampling of 2.4 pixels per
resolution element. The fibers in the SDSS spectroscopic survey
have a diameter of 3" on the sky, for our Type II AGNs sample at
redshifts $0.3 < z < 0.83$, the projected fiber aperture diameter
typically contains about 90\% of the total host galaxy light
(Kauffmann \& Heckman 2005), thus it is feasible to observe
significant stellar absorption features, which is the key point to
accurately measure the stellar velocity dispersion ($\sigma_{*}$)
in this paper. Since the SDSS spectra measure the light within a
fixed aperture size, the estimated velocity dispersions of more
distant galaxies could be affected by the rotation of stars at
larger physical radii than for nearby galaxies. In order to check
the rotation contribution to line broadening, we examine the
relation between the stellar velocity dispersion and the redshift,
and find no correlation at all, which suggests that the
line-broadening contribution from rotation of host galaxies is
very small and negligible. The main reason is that $\sigma_{*}$ is
derived through fitting heavy-element absorption lines (such as Mg
Ib$\lambda$ 5173, and CaII K $\lambda$ 3934, etc.), which are
mainly from the old stellar population in the bulge.

We first modelled the stellar contribution in the SDSS spectra of
type II AGNs through the modified version of the stellar
population synthesis code, STARLIGHT (version 2.0, Cid Fernandes
et al. 2001; Cid Fernandes et al. 2004a; Cid Fernandes et al.
2004b; Garcia-Rissman et al. 2005), which adopted the new stellar
library from Bruzual \& Charlot (2003). The code does a search for
the linear combination of Simple Stellar Populations (SSP) to
match a given observed spectrum $O_\lambda$. The model spectrum
$M_\lambda$ is:
\begin{equation}
M_\lambda(x,M_{\lambda_0},A_V,v_\star,\sigma_\star) =
M_{\lambda_0}
   \left[
   \sum_{j=1}^{N_\star} x_j b_{j,\lambda} r_\lambda
   \right]
   \otimes G(v_\star,\sigma_\star)
\end{equation}
where $b_{j,\lambda} \equiv L_\lambda^{SSP}(t_j,Z_j) /
L_{\lambda_0}^{SSP}(t_j,Z_j)$ is the spectrum of the $j^{\rm th}$
SSP normalized at $\lambda_0$, $r_\lambda \equiv 10^{-0.4
(A_\lambda - A_{\lambda_0})}$ is the reddening term, $x$ is the
population vector, $M_{\lambda_0}$ is the synthetic flux at the
normalization wavelength, and $G(v_\star,\sigma_\star)$ is the
line-of-sight stellar velocity distribution, modelled as a
Gaussian centered at velocity $v_\star$ and broadened by
$\sigma_\star$. The match between model and observed spectra is
calculated by $ \chi^2(x,M_{\lambda_0},A_V,v_\star,\sigma_\star) =
   \sum_{\lambda=1}^{N_\lambda}
   \left[
   \left(O_\lambda - M_\lambda \right) w_\lambda
   \right]^2
$, where the weighted spectrum $w_\lambda$ is defined as the
inverse of the noise in $O_\lambda$. For more detail, please refer
to Cid Fernandes et al. (2005).

Prior to the synthesis, the Galactic extinction  is computed by
using the extinction law of Cardelli, Clayton \& Mathis (1989) and
the $A_V$ value is taken from Schlegel, Finkbeiner \& Davis (1998)
as listed in the NASA/IPAC Extragalactic Database (NED). The
spectra are transformed into the rest frame defined by the
redshift given in their SDSS FITS header. The spectrum is
normalized at 4020$\AA$ and the signal-to-noise ratio is measured
in the S/N window between 4730 and 4780 $\AA$. Masks of $20-30$
$\AA$ around obvious emission lines are constructed for each
object individually. Because the redshift coverage of this type II
AGNs sample, we focus on the strongest stellar  absorption
features of Ca II K and the G-band, which are less affected by
nearby emission lines. An additional $f_{\nu} \sim \nu ^{-1.5}$
power-law component (Watanabe et al. 2003) is used to account for
the contribution from the scattered AGNs continuum emission, a
traditional ingredient for modeling Seyfert galaxies since Koski
(1978). Finally we check visually our spectral fitting results one
by one.

For our sample, the S/N in the S/N window varies between 0.3 and
21.5. The  fitting results for high S/N objects are usually better
than those for low S/N  ones.  After inspecting the fitting
results, we find that the fitting goodness  (chi-square value)
depends not only on the S/N ($>5$ , in the given S/N spectral
window), but also on the absorption lines equivalent widths (EW of
Ca II K line $>1.5\AA$). At last we select 33 type II AGNs, which
had shown significant stellar absorption features and are well
fitted to derive reliable measurements of stellar velocity
dispersion.

In order to check  whether this sub-sample is representative of
the total sample of Zakamsa et al. (2003) with respect to the [O
III]$\lambda$5007 luminosity ($L_{\rm [O III]}$), we plot the
histograms of the $L_{\rm [OIII]}$ distribution for the sub-sample
and total-sample (see Figure 1). Then T-test shows that these two
populations are drawn from the same parent population with a
possibility of 0.95. $L_{\rm [O III]}$ is directly adopted from
Table 1 in Zakamsa et al. (2003), the adopted raw $L_{\rm [OIII]}$
values and associated quantities are lower limits. Using the
$L_{\rm [O III]}$ criterion of $3\times 10^{8} \Lsolar$(the
logarithm is $\sim 8.48$), which corresponds to the intrinsic
absolute magnitude $M_{B} < -23$ (Zakamsa et al. 2003), 20 objects
can then be classified as type II quasars. Fig. 2 shows a fitting
example of for SDSS J150117.96+545518.2 with S/N=20.5. The final
results are presented in Table 1. All the fittings for 33 type II
AGNs are appended in the Appendix.

After subtracting the synthetic stellar components and the AGNs
continuum, we obtain the clean pure emission-line spectra  as
shown in the top panel of Fig. 2, where we can  analyze the pure
emission-line profiles in detail by  using the multi-component
spectral fitting task SPECFIT (Kriss 1994) in the IRAF-STSDAS
package {\footnote {IRAF is distributed by the National Optical
Astronomy Observatory, which is operated by the Association of
Universities for Research in Astronomy, Inc., under cooperative
agreement with the National Science Foundation.}}. Because of the
asymmetric profiles of the [O III]$\lambda\lambda$4959, 5007
lines, two sets of two-gaussian profiles are used in order to
remove properly the asymmetric blue/red wings of the [O III] line.
We take the same linewidth for each component, and fix the flux
ratio of [O III]$\lambda$4959 to [O III]$\lambda$5007 to be 1:3.
For three objects, we can't fit the [O III] lines for their
irregular [O III] lines (see Table 1). For the [O
II]$\lambda\lambda$ 3727, 3729 lines, we use two-gaussian profiles
and  fix their wavelength separation to the laboratory value; the
ratio of the line intensities is allowed to vary during the
fitting. The decomposition for [O II] lines is more difficult
because of relatively low S/N and that the expected line widths
are comparable to the pair separation (2.4\AA). For objects with
available H$\alpha$, we also used one-gaussian profile to fit NII
$\lambda \lambda$ 6548,6583 and H$\alpha$ 6563 lines. And
H$\alpha$ and H$\alpha$ luminosities would be used to do the
extinction correction. For more details, please refer to Bian,
Yuan \& Zhao (2005, 2006). Our sample fitting for SDSS
J150117.96+545518.2 is showed in Fig. 3.

\section{Results and discussion}
\subsection{The uncertainties of the stellar velocity dispersion
($\sigma_{*}$) and the gaseous velocity dispersion ($\sigma_{g}$)}

The derived stellar velocity dispersion was corrected by the
instrumental resolutions of both the SDSS spectra and the STELIB
library. Cid Fernandes et al. (2004a) have used their stellar
population synthesis method to study a sample of 79 nearby
galaxies observable from the southern hemisphere, of which 65 are
Seyfert 2 galaxies. The S/N in the S/N window varies between 10
and 67. They compared their $\sigma_{*}$ with values from the
literature and found the agreement is good. They estimated that
the uncertainty in $\sigma_{*}$ is typically about 20 \kms.
Recently, Cid Fernandes et al. (2005) applied their synthesis
method to a larger sample of 50362 galaxies from the SDSS Data
Release 2 (DR2). Their derived $\sigma_{*}$ is consistent very
well with that of the MPA/JHU group (Kauffmann et al. 2003). The
median of the difference between the two estimates is just 9 \kms.
We have carefully checked our synthesis fitting result one by one
and picked out 33 type II AGNs that are well fitted and the
stellar velocity dispersion ($\sigma_{*}$) are reliably derived.
The spectral S/N for these objects are in the range of 5 to 21.5,
most of which larger than 10, thus the typical uncertainty in
$\sigma_{*}$ should be around 20 \kms.

Recently, Greene \& Ho (2006a, 2006b) used the direct-fitting
method (Barth et al. 2002) to study the the systematic biases of
$\sigma_{*}$ from different regions around Ca II triplet, Mg Ib
triplet, and CaII H+K stellar absorption features, which are
introduced by both template mismatch and contamination from AGNs.
They argue that the Ca II triplet provides the most reliable
measurements of $\sigma_{*}$ and there is a systematic offset
between $\sigma_{\rm Ca K}$ and $\sigma_{*}$ derived from other
spectral regions. For our higher-redhsift sample and the SDSS
wavelength coverage 3800-9200\AA, it is impossible to measure
$\sigma_{*}$ from Ca II triplet. Therefore, for higher-redhsift
type II AGNs, new observations around Ca II triplet are necessary
in the future. Here we used the following formula to obtain the
corrected velocity dispersion $\sigma_{*}^c$ (Greene \& Ho 2006b),
\begin{equation}
\sigma_*^c = (1.40\pm0.04)\sigma_{*} - (71\pm5) .
\end{equation}
For three objects, $\sigma_*$ is near the instrumental resolution
and the corrected $\sigma_*^c$ is  unreliable. These objects are
excluded from further analysis, and denoted as $\dag$ in Table 1.

The gaseous velocity dispersion ($\sigma_{\rm g}$) is obtained
from full width half maximum (FWHM) of emission lines by assuming
the Gaussian profile: $\rm \sigma_{\rm [O II]}^{g}=FWHM[O
II]/2.35$ and $\rm \sigma_{\rm [O III]}^{g}=FWHM[O III]/2.35$.
Taking into account the SDSS spectrum resolution, the intrinsic
$\sigma_{g}$ derived from $\rm FWHM([O III])$ may be
instrumentally broadened. The intrinsic $\sigma_{\rm g}$ value can
be approximated by $\sigma_{\rm g}=(\sigma_{\rm
obs}^2-(\sigma_{\rm inst}/(1+z))^2)^{1/2}$, where z is the
redshift. For the spectra from SDSS, the mean values of
$\sigma_{\rm inst}$ are 74 \kms (the logarithm is 1.87 dex) for [O
II], and 60 \kms (the logarithm is 1.78 dex) for [O III] (Greene
\& Ho 2005), respectively. The results of $\sigma_{\rm g}$ are
listed in Table 1 (columns 7 and 8). After removing the effect of
instrumental broadening, some objects become unresolved (see Table
1). Measurements of $\sigma_{\rm g}$ below the resolution limit
(74 and 60 \kms for [O II] and [O III], respectively) are not
reliable. The error of $\sigma_{\rm g}$ is derived from the
measurement error of the linewidth. For the linewidth, the typical
error is about 10 per cent. However, the systematic effects are
neglected, e.g. the uncertainties of the continuum subtraction and
component decomposition (Bian, Yuan \& Zhao 2005).

\subsection{The extinction correction of the [O III]$\lambda$5007 luminosity}
The [O III] line has the advantage of being strong and easy to
detect in AGNs. The [O III]$\lambda$5007 line is usually used as a
tracer of AGNs activity (e.g. Heckman et al., 2004; Greene \& Ho
2005). However, the [O III]$\lambda$5007 luminosity is subject to
extinction by interstellar dust in the host galaxy and in our
Galaxy. Using SDSS data, Kauffmann et al (2003) found that for
type II AGNs at $z<0.3$ the mean value of the ratio of the
corrected luminosity to the observational luminosity ($log
(\frac{L^{c}_{[OIII]}}{L_{[O III]}})$, the extinction factor at
5007$\AA$) is about 0.85 dex and suggested that the extinction
effect should be huge for type II quasars. The extinction
correction is usually corrected by using the Balmer decrement,
which is regarded as the best approximation. For 18 out of these
33 objects with available H$\alpha$ measurements in the
stellar-light subtracted spectra, we calculated the extinction
correction of the [O III]$\lambda$5007 luminosity, from the
relation (Bassani et al. 1999)
\begin{equation}
L_{\rm [O III]}^{\rm c} = L_{\rm [O III]} \rm
[\frac{(H_\alpha/H_\beta)_{obs}}{(H_\alpha/H_\beta)_{0}}]^{2.94}
\end{equation}
where an intrinsic Balmer decrement $\rm (H_\alpha/H_\beta)_{0} =
3.0$ is adopted (Gu \& Huang 2002). The range of the extinction
factors at [OIII] is between -0.24 dex and 1.25 dex.  The
extinction-corrected [O III]$\lambda$5007 luminosity is listed in
Col. 11 in Table 1. If we used the $L_{\rm [O III]}$ criterion of
$3\times 10^{8} \Lsolar$ (Zakamsa et al. 2003), 14 out of these 18
objects can then be classified as type II quasars.

Using the SDSS AGN catalogue at MPA/JHU (Kauffmann et al
2003)\footnote{http://www.mpa-garching.mpg.de/SDSS/DR4/index.html},
we obtained the extinction factor at 5007$\AA$, $log
(\frac{L^{c}_{[OIII]}}{L_{[O III}})$.  We tried to obtain a
relation between $\frac{L^{c}_{[OIII]}}{L_{[O III}}$ and other
observational parameters, such as $\sigma_*$, $L_{[O III}]$,
$L_{H\beta}$. However, no correlation is found. More careful work
is required to deal with this problem in the future.

\subsection{The SMBHs masses and Eddington ratios in type II quasars}
Using the reverberation mapping method or the empirical
size-luminosity relation, it is impossible to estimate the SMBHs
masses in type II quasars for the lack of emission lines from
BLRs. Here we use the formulae to derive the SMBHs masses in type
II quasars from the stellar velocity dispersion $\sigma_{*}$
(Tremaine et al. 2002),  which is $M_{\rm BH} (\Msolar)=
10^{8.13}(\sigma_{*}^{c}/(200 \ \rm km s^{-1}))^{4.02} $.

We calculate the Eddington ratio, $L_{\rm bol}/L_{\rm Edd}$. We
use the extinction-corrected [O III] luminosity ($L^{c}_{[O III}$)
as a surrogate for the AGNs luminosity (Heckman et al., 2004;
Greene \& Ho 2005), $\rm L_{\rm bol}=3500L^{\rm c}_{\rm [O III]}$,
to calculate the bolometric luminosity $L_{\rm bol}$, and $\rm
L_{\rm Edd}=1.26\times 10^{38}M_{\rm BH}/\Msolar~ergs~ s^{-1}$.
Objects with no extinction corrected [O III] luminosity are
discarded in the analysis relative to the Eddington ratio. The
results of SMBHs masses and Eddington ratios in type II AGNs are
also presented in Table 1 (columns 13 and 14). We also calculate
the SMBHs masses and Eddington ratios for the lower-redshifts type
II AGNs at $0.02 < z < 0.3$ presented by Kauffmann et al. (2003).
For the typical uncertainties of 20\kms for $\sigma_{*}=200$ \kms,
the errors of $log~ \sigma_{*}$ would be about 0.05 dex,
corresponding 0.17 dex for $log M_{\rm BH}$, and almost the same
for $L_{\rm bol}/L_{\rm Edd}$ (Bian, Yuan, \& Zhao 2005). Here we
don't consider the cosmology evolution of $M_{\rm BH}-\sigma_{*}$
relation (e.g. Woo et al. 2006), which is a question open to
debate.

In Fig. 4, we compare the distribution of SMBHs masses and
Eddington ratios in the lower-redshifts sample and the
higher-redshifts sample. It is found that the type II AGNs at
higher-redshifts have higher SMBHs masses and higher Eddington
ratios.

It is well known that the Eddington ratio is an important
parameter to describe the accretion process in AGNs. The [O III]
luminosity is usually used as a surrogate for the AGNs luminosity
(Heckman et al. 2004 and the reference therein). In Fig. 5, we
plot the [O III] luminosity and the corrected [O III] luminosity
versus the SMBHs masses. Using a least-squares regression, we
derive the correlation between $M_{\rm BH}$ and $L_{\rm [O III]}$
to be: $\rm log(L_{[O
III]}/L_{\odot})=(6.83\pm0.65)+(0.22\pm0.08)log(M_{BH}
/M_{\odot})$. The correlation coefficient R is 0.45, with a
probability of $P_{null} =0.012$ for rejecting the null hypothesis
of no correlation. However, no correlation is found between
$M_{\rm BH}$ and $L^{c}_{\rm [O III]}$. No correlation is found
between the Eddington ratio and the [O III] luminosity or the
corrected [O III] luminosity. From the photoionization model, the
strength of [O III] is controlled by the NLRs covering factor,
density, and ionization parameter (e.g. Baskin \& Laro 2005). The
relation between the [O III] luminosity and the AGNs bolometric
luminosity is still a question to debate.

\subsection{The $\sigma_{\rm g}-\sigma_{*}$ relation}
The existence of a good correlation between stellar velocity
dispersion ($\sigma_{*}$) and ionized gas velocity dispersion
($\sigma_{\rm g}$) (e.g. Nelson \& Whittle 1996) suggests that the
gaseous kinematics of NLRs in Seyfert galaxies be primarily
dominated by the bulge gravitational potential, which is further
confirmed by Nelson \& Whittle (1996). Most recently, Greene \& Ho
(2005) have investigated a large and homogeneous sample of
lower-redshift (0.02 $<$ z $<$ 0.3) Type II AGNs from the SDSS,
and found that $\sigma_{\rm g}$ traces $\sigma_{*}$. Though the
gas kinematics of NLRs are governed by the gravitational potential
of the bulges, they also find that the accretion rate plays an
important secondary role.

Following Greene \& Ho (2005), we study the $\sigma_{\rm
g}-\sigma_{*}^{c}$ relation for 33 Type II AGNs at redshifts 0.3
$<$ z $<$ 0.83 after measuring  the gas velocity dispersion
($\sigma_{\rm g}$) from the narrow emission lines from NLRs, and
the stellar velocity dispersion ($\sigma_{*}^{c}$) from the CaII
H+K, G-band absorption feature,  which is shown in Fig. 2. In Fig.
6, we showed the relation between $\sigma_{g}$ and $\sigma_{*}^c$.
The correlation coefficients are 0.19, 0.27, 0.70, 0.54 for figurs
in Fig. 6, from left to right and from top to bottom. The relation
between $\sigma_{g}$ and $\sigma_{*}^c$ becomes much weaker at
higher redshifts. We also found that this poor correlation is not
due to the S/N. In order to qualify the comparison between
$\sigma_{\rm g}$ and $\sigma_{*}^{c}$, we calculate the
distribution of $<\sigma_{\rm g}/\sigma_{*}^{c}>$. The value is
$1.24\pm0.76$ for the the core component of [O III] line,
$1.20\pm0.96$ for the [O II] line. These suggest that $\sigma_{\rm
g}$ of the the core component of [O III] and [O II] lines can
trace $\sigma_{*}^{c}$ within 0.09 and 0.08 dex in the logarithm
of $\sigma_{*}^{c}$, respectively, and that the high-ionization [O
III] line traces $\sigma_{*}^{c}$ as well as the low-ionization [O
II] line. If we use the line width of [O III] core component to
estimate the SMBHs masses from
the Tremaine's M$_{\rm BH} - \sigma$ relation, we would
overestimate SMBHs masses by about 0.38 dex.

For comparison, Greene \& Ho (2005) derived the distribution of
$<\sigma_{\rm g}/\sigma_*>$ of lower-redshift type II AGNs with
$0.02 < z < 0.3$, which are $1.34\pm0.66$ for the [O III] line,
$1.00\pm0.35$ for the [O III] core line,  and $1.13\pm0.38$ for
the [O II] line. We also calculate the the distribution of
$<\sigma_{\rm g}/\sigma_{*}^{c}>$ for the sample of Seyfert
galaxies (Nelson \& Whittle 1996); the value is $1.15\pm0.68$,
suggesting that $\sigma_{\rm g}$ of the [O III] line can trace
$\sigma_{*}^{c}$ within 0.06 dex in the logarithm of
$\sigma_{*}^{c}$ . Our
 results are thus consistent with theirs.

To a first-order approximation, the line widths of the core
component of [O III] and [O II] for both low-redshift and
high-redshift Type II AGNs are primarily controlled by the
gravitational potential of the bulges of host galaxies, and can
approximately trace the stellar velocity dispersion. As we know,
the errors in virial SMBHs masses derived from galaxy dynamics or
size-luminosity relation is about ~0.5 dex (e.g. Magorrian et al.
1998; Kaspi et al. 2000; Bian \& Zhao 2004a, 2004b). We also can
use the line-width of the core component of [O III] or [O II] to
estimate the black hole mass.

In order to  find the secondary effect of the line broadening in
gas lines from NLRs for our higher-redshift narrow-line AGNs, we
studied the relation between the deviation of $\sigma_{\rm g}$
from $\sigma_{*}^{c}$ ($\Delta \sigma=log\sigma_{\rm
g}-log\sigma_{*}^{c}$) and the Eddington ratio ($L_{\rm bol}
/L_{\rm Edd}$) (Greene \& Ho 2005). Using the least-square
regression, we find a median strong correlation between these two
dimensionless parameters. For the [O III] line, the relation is:
$\Delta \sigma=(-0.035\pm0.057)+(0.202\pm0.058)log(L_{\rm bol}
/L_{\rm Edd})$, R=0.71, $P_{\rm null} = 0.00441$. For the [O II]
line, the relation is $\Delta
\sigma=(0.047\pm0.058)+(0.179\pm0.059)log(L_{\rm bol} /L_{\rm
Edd})$, R=0.65, $P_{\rm null} =0.0091$ (see Fig. 7). We also find
the comparable correlation between $\Delta \sigma$ and $M_{\rm
BH}$. These results confirm that the nuclei accretion process
and/or nuclei SMBHs would affect the linewidth of gas lines from
NLRs, although the primary driver is the gravitational potential
of the bulge.

\section{Conclusion}
The stellar population synthesis code is used to model the stellar
contribution for a sample of 209 type II AGNs at redshifts
$0.3<z<0.83$ from SDSS. According to the $L_{\rm [O III]}$
criterion of $3\times 10^{8} \Lsolar$, 20  can be classified as
type II quasars. The main conclusions can be summarized as
follows.

\begin{itemize}

\item{The reliable $\sigma_{*}$ are measured for 33 type II AGNs
with significant stellar absorption features. We use the formula
of Greene \& Ho to obtain the corrected stellar velocity
dispersions ($\sigma_{*}^c$ and SMBHs masses are calculated from
the $M_{\rm BH}-\sigma_{*}^{c}$ relation. A median strong relation
between the [O III] luminosity and the SMBH mass is found
(although no correlation between the extinction-corrected [O III]
luminosity and the SMBH mass); no correlation is found between the
Eddington ratio and the [O III] luminosity or the
extinction-corrected [O III] luminosity.}

\item{The gas velocity dispersion ($\sigma_{g}$) in NLRs is
measured using three sets of two-gaussian profiles to fit [O
III]$\lambda\lambda$4959, 5007 and [O II]$\lambda\lambda$3727,
3729) in these 33 stellar-light subtracted spectra. We find that
the relation between $\sigma_{g}$ and $\sigma_{*}^c$ becomes much
weaker at higher redshifts.}

\item{The distribution of $<\sigma_{g}/\sigma_{*}^c>$ is
$1.24\pm0.76$ for the core [O III] line and $1.20\pm0.96$ for the
[O II] line, which suggests that $\sigma_{g}$ can trace
$\sigma_{*}^c$ within about 0.1 dex in the logarithm of
$\sigma_{*}^c$. The deviation of $\sigma_{g}$ from $\sigma_{*}^c$
is correlated with the Eddington ratio.}

\end{itemize}

\section*{ACKNOWLEDGMENTS}
We thank Luis C. Ho for his helpful comments. We thank the
anonymous referee for his/her comments and instructive
suggestions, which significantly improved our work. We are
grateful to Dr. Helmut Abt for checking our manuscript. This work
has been supported by the NSFC (Nos. 10403005 and 10473005) and
the Science-Technology Key Foundation from Education Department of
China (No. 206053). QGU would like to acknowledge the financial
supports from China Scholarship Council (CSC) and the National
Natural Science Foundation of China under grants 10103001 and
10221001. Funding for the creation and distribution of the SDSS
Archive has been provided by the Alfred P. Sloan Foundation, the
Participating Institutions, NASA, the National Science Foundation,
the US Department of Energy, the Japanese Monbukagakusho, and the
Max Planck Society. The SDSS Web site is http:// www.sdss.org/.
This research has made use of the NASA/IPAC Extragalactic
Database, which is operated by the Jet Propulsion Laboratory at
Caltech, under contract with NASA.

\begin{table*}
\begin{center}
\begin{tiny}
\begin{tabular}{llllllllllllllllllllllll}
\hline \hline

name   &z& $\sigma_*$ & $[O III]$   & $[O II]$ & $\sigma_*^c$ &  $\sigma_g^{[O III]}$ &  $\sigma_g^{[O II]}$  &$L_{[O III]}$& $L_{[O II]}$&$L^{c}_{[O III]}$&S/N  & $M_{bh}$ & $\frac{L_{bol}}{L_{Edd}}$ \\
      (1) & (2) &(3) & (4) & (5) & (6)&(7)& (8) & (9)& (10)& (11) & (12)& (13)&(14)\\
\hline
 SDSS J113344.02+613455.7& 0.426 &273 & $327 \pm 98    $&    $ 526  \pm  71 $ &2.49 & $2.13^{+0.12}_{-0.18}$&       $ 2.34 ^{+ 0.06}_{-0.07}$&           8.82 & 8.52  &--  &  21.5  &  8.90&   --      \\
 SDSS J150117.96+545518.2& 0.338 &289 & $456 \pm 48    $&    $ 436  \pm  45 $ &2.52 & $2.28^{+0.05}_{-0.05}$&       $ 2.25 ^{+ 0.04}_{-0.05}$&           9.06 & 8.42  &8.97&  20.5  &  9.02&   -0.81   \\
 SDSS J094820.38+582526.4& 0.324 &126 & $470 \pm 67    $&    $ 750  \pm  56 $ &2.02 & $2.29^{+0.06}_{-0.07}$&       $ 2.50 ^{+ 0.03}_{-0.03}$&           7.89 & 8.25  &9.08&  20.2  &  7.01&   0.54    \\
 SDSS J002531.45-104022.2& 0.303 &82  & $485 \pm 60    $&    $ 529  \pm  55 $ &1.64$\dag$ & $2.31^{+0.05}_{-0.06}$&       $ 2.34 ^{+ 0.04}_{-0.05}$&    8.73   &9.02& 8.57  &  16.7  &  5.48&  2.53      \\
 SDSS J083945.98+384319.0& 0.424 &196 & $597 \pm 51    $&    $ 723  \pm  58 $ &2.31 & $2.40^{+0.04}_{-0.04}$&       $ 2.48 ^{+ 0.03}_{-0.04}$&           8.70 & 8.40  &--  &  16.1  &  8.16&   --      \\
 SDSS J104505.39+561118.3& 0.428 &204 & $413 \pm 116   $&    $ 505  \pm  47 $ &2.33 & $2.23^{+0.11}_{-0.15}$&       $ 2.32 ^{+ 0.04}_{-0.04}$&           9.08 & 8.86  &--  &  15.8  &  8.25&   --      \\
 SDSS J090414.10-002144.9& 0.353 &287 & $409 \pm 49    $&    $ 297  \pm  32 $ &2.52 & $2.23^{+0.05}_{-0.06}$&       $ 2.06 ^{+ 0.05}_{-0.06}$&           8.93 & 8.10  &9.03&  15.4  &  9.01&   -0.91   \\
 SDSS J143027.66-005614.8& 0.318 &157 & $367 \pm 118   $&    $ 453  \pm  59 $ &2.17 & $2.18^{+0.13}_{-0.19}$&       $ 2.27 ^{+ 0.06}_{-0.06}$&           8.36 & 7.87  &9.23&  14.8  &  7.61&   0.48    \\
 SDSS J133633.65-003936.4& 0.416 &153 & $486 \pm 181   $&    $ 470  \pm  94 $ &2.16 & $2.31^{+0.14}_{-0.22}$&       $ 2.29 ^{+ 0.08}_{-0.10}$&           8.64 & 7.84  &--  &  14.8  &  7.55&   --      \\
 SDSS J100459.41+030202.0& 0.469 &130 & $326 \pm 13    $&    $ 369  \pm  57 $ &2.05 & $2.13^{+0.02}_{-0.02}$&       $ 2.17 ^{+ 0.07}_{-0.08}$&           8.77 & 8.25  &--  &  14.8  &  7.10&   --       \\
 SDSS J020234.55-093921.8& 0.302 &211 & $466 \pm 41    $&    $ 524  \pm  68 $ &2.35 & $2.29^{+0.04}_{-0.04}$&       $ 2.33 ^{+ 0.05}_{-0.06}$&           8.12 & 7.80  &8.52&  14.0  &  8.33&   -1.09   \\
 SDSS J094209.00+570019.7& 0.35  &102 & $597 \pm 24    $&    $ 343  \pm  58 $ &1.86 & $2.40^{+0.02}_{-0.02}$&       $ 2.13 ^{+ 0.07}_{-0.09}$&           8.31 & 8.34  &8.82&  12.8  &  6.34&   1.33     \\
 SDSS J154337.81-004420.0& 0.311 &207 & $200 \pm 39    $&    $ 482  \pm  60 $ &2.34 & $1.87^{+0.10}_{-0.14}$&       $ 2.29 ^{+ 0.05}_{-0.06}$&           8.40 & 7.76  &8.33&  11.9  &  8.29&   -1.01   \\
 SDSS J033606.70-000754.7& 0.431 &102 & $774 \pm 52    $&    $ 562  \pm  82 $ &1.86 & $2.51^{+0.03}_{-0.03}$&       $ 2.37 ^{+ 0.06}_{-0.07}$&           8.71 & 7.98  &--  &  11.8  &  6.34&   --       \\
 SDSS J075607.15+461411.5& 0.593 &316 & $322 \pm 60    $&    $ 319  \pm  279$ &2.57 & $2.12^{+0.08}_{-0.10}$&       $ 2.11 ^{+ 0.28}_{-1.20}$&           9.01 & 8.18  &--  &  10.8  &  9.21&   --      \\
 SDSS J084041.05+383819.9& 0.313 &163 & $720 \pm 48    $&    $ 650  \pm  94 $ &2.20 & $2.48^{+0.03}_{-0.03}$&       $ 2.43 ^{+ 0.06}_{-0.07}$&           8.62 & 8.03  &9.48&  10.8  &  7.71&   0.45    \\
 SDSS J143928.23+001538.0& 0.339 &184 & $569 \pm 56    $&    $ 652  \pm  58 $ &2.27 & $2.38^{+0.04}_{-0.05}$&       $ 2.43 ^{+ 0.04}_{-0.04}$&           8.08 & 7.99  &9.22&  10.8  &  8.01&   -0.29   \\
 SDSS J231755.35+145349.3& 0.311 &48  & $229 \pm 6     $&    $ 264  \pm  54 $ &----  $\dag$ & $1.94^{+0.01}_{-0.01}$&       $ 1.99 ^{+ 0.09}_{-0.12}$&   8.10 & 7.66  &8.39&  10.7  &  -   &   --       \\
 SDSS J121856.41+611922.6& 0.369 &100 & $182 \pm 28    $&    $ 270  \pm  27 $ &1.84 & $1.82^{+0.08}_{-0.11}$&       $ 2.01 ^{+ 0.05}_{-0.05}$&           8.38 & 8.07  &8.60&  10.7  &  6.27&   1.33     \\
 SDSS J075920.21+351903.4& 0.328 &214 & ---       &    $ 569  \pm  88 $ &2.36 &  ---                    &       $ 2.37 ^{+ 0.06}_{-0.08}$&               7.59 & 7.88  &8.84&  10.4  &  8.36&   -0.91   \\
 SDSS J132529.32+592424.9& 0.429 &328 & $443 \pm 61    $&    $ 441  \pm  42 $ &2.59 & $2.27^{+0.06}_{-0.07}$&       $ 2.26 ^{+ 0.04}_{-0.05}$&           8.89 & 8.37  &--  &  10.2  &  9.29&   --      \\
 SDSS J090626.80+033310.7& 0.363 &127 & $191 \pm 36    $&    $ 271  \pm  70 $ &2.03 & $1.85^{+0.10}_{-0.13}$&       $ 2.01 ^{+ 0.11}_{-0.16}$&           7.73 & 7.17  &--  &  10.2  &  7.03&   --      \\
 SDSS J031012.82-010822.6& 0.303 &70  & $200 \pm 51    $&    $ 242  \pm  68 $ &1.43$\dag$ & $1.87^{+0.12}_{-0.20}$&       $ 1.93 ^{+ 0.12}_{-0.19}$&     8.06 & 7.73  &8.40&  10.1  &  4.63&   2.74     \\
 SDSS J092318.04+010144.7& 0.386 &230 & $694 \pm 64    $&    $ 406  \pm  68 $ &2.40 & $2.47^{+0.04}_{-0.04}$&       $ 2.22 ^{+ 0.07}_{-0.09}$&           8.94 & 7.94  &9.31&  9.9   &  8.53&   -0.34   \\
 SDSS J081507.42+430427.1& 0.510 &335 & $860 \pm 35    $&    $ 1070 \pm  76 $ &2.60 & $2.56^{+0.02}_{-0.02}$&       $ 2.66 ^{+ 0.03}_{-0.03}$&           9.57 & 9.02  &--  &  9.8   &  9.33&   --      \\
 SDSS J033248.49-001012.3& 0.310 &339 & $627 \pm 117   $&    $ 672  \pm  55 $ &2.61 & $2.42^{+0.08}_{-0.09}$&       $ 2.45 ^{+ 0.03}_{-0.04}$&           8.50 & 7.97  &8.26&  9.6   &  9.36&   -2.23   \\
 SDSS J092014.11+453157.3& 0.402 &174 & $666 \pm 50    $&    $ 447  \pm  70 $ &2.24 & $2.45^{+0.03}_{-0.03}$&       $ 2.26 ^{+ 0.07}_{-0.08}$&           9.04 & 8.43  &--  &  9.6   &  7.87&   --       \\
 SDSS J214415.61+125503.0& 0.390 &182 & $426 \pm 43    $&    $ 614  \pm  124$ &2.26 & $2.25^{+0.04}_{-0.05}$&       $ 2.41 ^{+ 0.08}_{-0.10}$&           8.14 & 7.81  &--  &  8.6   &  7.98&   --      \\
 SDSS J153943.73+514221.0& 0.585 &240 & $427 \pm 43    $&    $ 349  \pm  79 $ &2.42 & $2.25^{+0.04}_{-0.05}$&       $ 2.15 ^{+ 0.09}_{-0.12}$&           8.47 & 8.37  &--  &   8.5  &  8.62&   --      \\
 SDSS J133735.01-012815.6& 0.329 &254 & $331 \pm 130   $&    $ 265  \pm  264$ &2.45 & $2.13^{+0.15}_{-0.26}$&       $ 1.99 ^{+ 0.33}_{-1.20}$&           8.72 & 8.28  &9.56&  6.8   &  8.75&   -0.49   \\
 SDSS J165627.28+351401.7& 0.679 &183 & ---      &    $ 573  \pm  66 $ &2.27 &  ---                    &       $ 2.38 ^{+ 0.05}_{-0.05}$&                8.57 & 7.91  &--  &  6.7   &  8.00&   --      \\
 SDSS J101237.29+023554.3& 0.720 &139 & ---       &    $ 799  \pm  98 $ &2.09 &  ---                   &       $ 2.53 ^{+ 0.05}_{-0.06}$&                8.22 & 8.19  &--  &  5.1   &  7.29&   --      \\
 SDSS J094836.03+002104.5& 0.324 &130 & $555 \pm 116   $&    $ 515  \pm  40 $ &2.05 & $2.37^{+0.08}_{-0.11}$&       $ 2.33 ^{+ 0.03}_{-0.04}$&           8.52 & 7.90  &8.81&  4.9   &  7.10&   0.62    \\                                                                                                                                                                                                       \hline
\end{tabular}
\end{tiny}
\end{center}
\caption{Results for type II quasars, sorted by the value of S/N.
The columns are as follows: (1) name; (2) redshift; (3)
uncorrected stellar velocity dispersion in units of \kms; (4) FWHM
of [O III]$\lambda$ 5007 \AA ~in units of \kms; (5) FWHM of [O
II]$\lambda$ 3727 \AA ~in units of \kms; (6) log of corrected
stellar velocity dispersion by equation 2 (in units of \kms); (7)
log of the [O III] $\sigma$ in units of \kms; (8) log of the [O
II] $\sigma$ in units of \kms; (9) log of [O III] luminosity in
unit of \Lsolar; (10) log of [O II] luminosity in unit of \Lsolar;
(11) log of extinction-corrected [O II] luminosity in unit of
\Lsolar; (12) signal-to-noise ratio (S/N); (13) log of the black
hole mass in units of solar mass; (14) log of the Eddington ratio.
$\dag$: $\sigma_*$ is near the instrumental resolution and the
corrected $\sigma_*^c$ is unreliable, which is excluded from
future analysis.}
\end{table*}

\begin{figure*}
\begin{center}
\includegraphics[width=16cm,height=10cm]{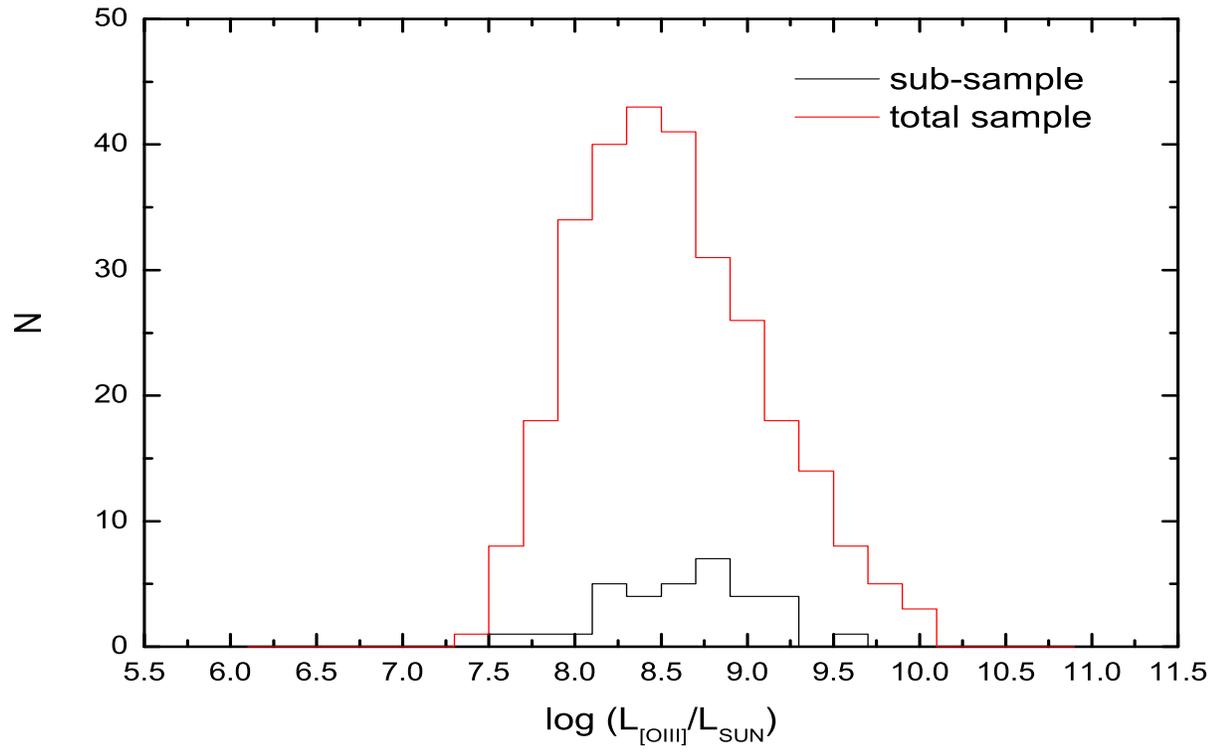}
\caption{The distribution of $L_{\rm [O III]}$ for total sample
and the sub-sample.}
\end{center}
\end{figure*}

\begin{figure*}
\begin{center}
\includegraphics[width=16cm,height=10cm]{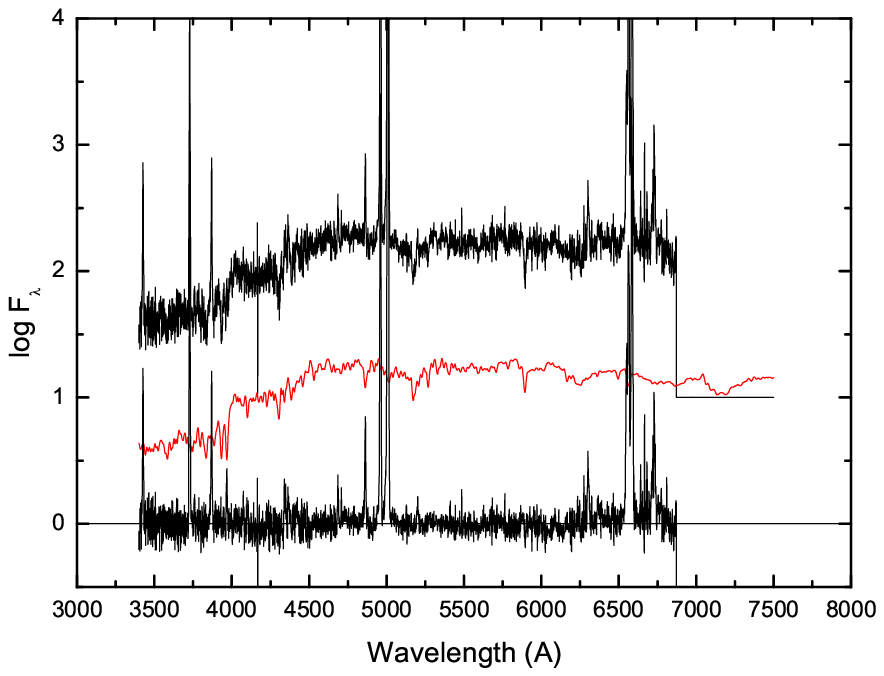}
\includegraphics[width=6cm,height=8cm,angle=-90]{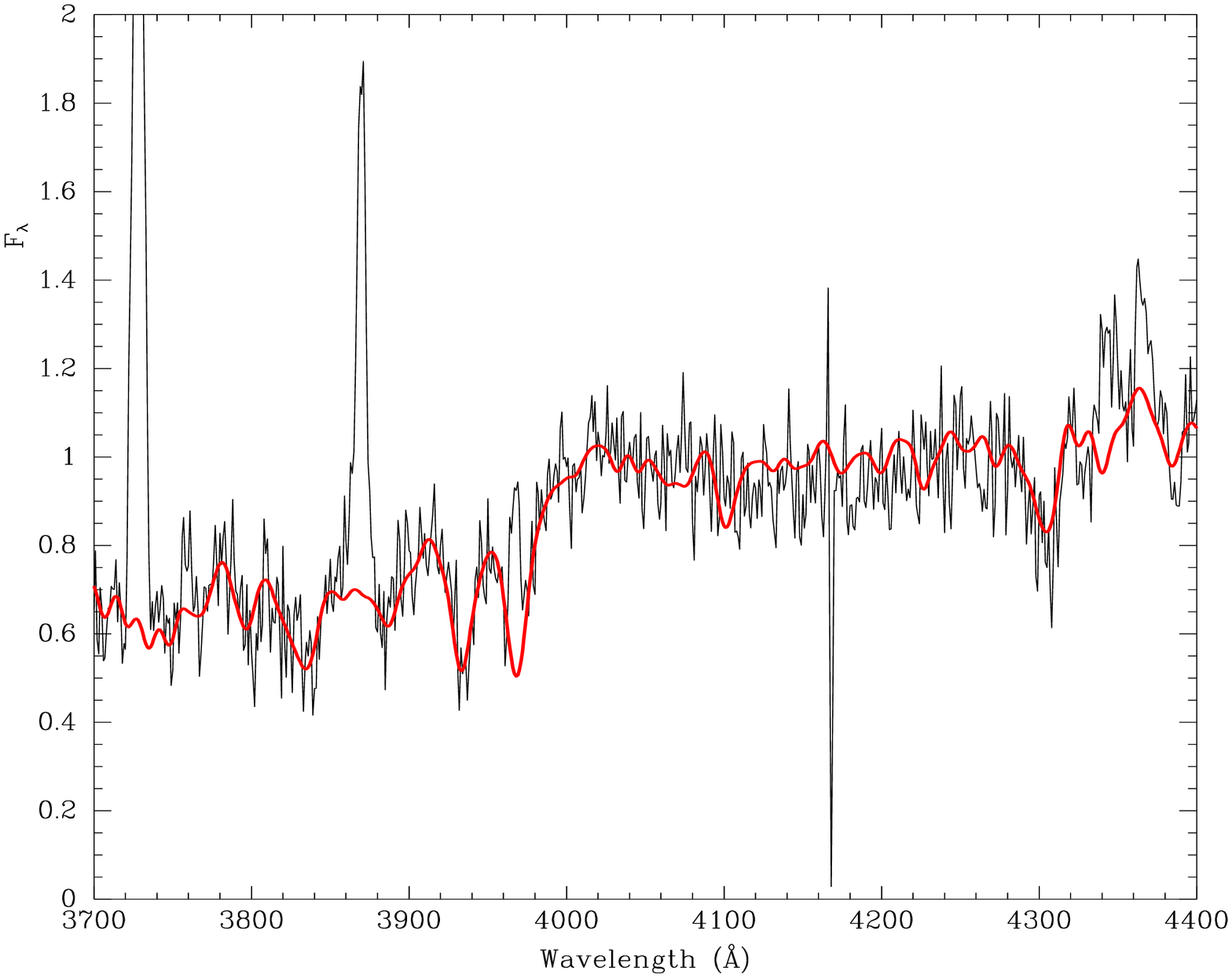}
\includegraphics[width=6cm,height=8cm,angle=-90]{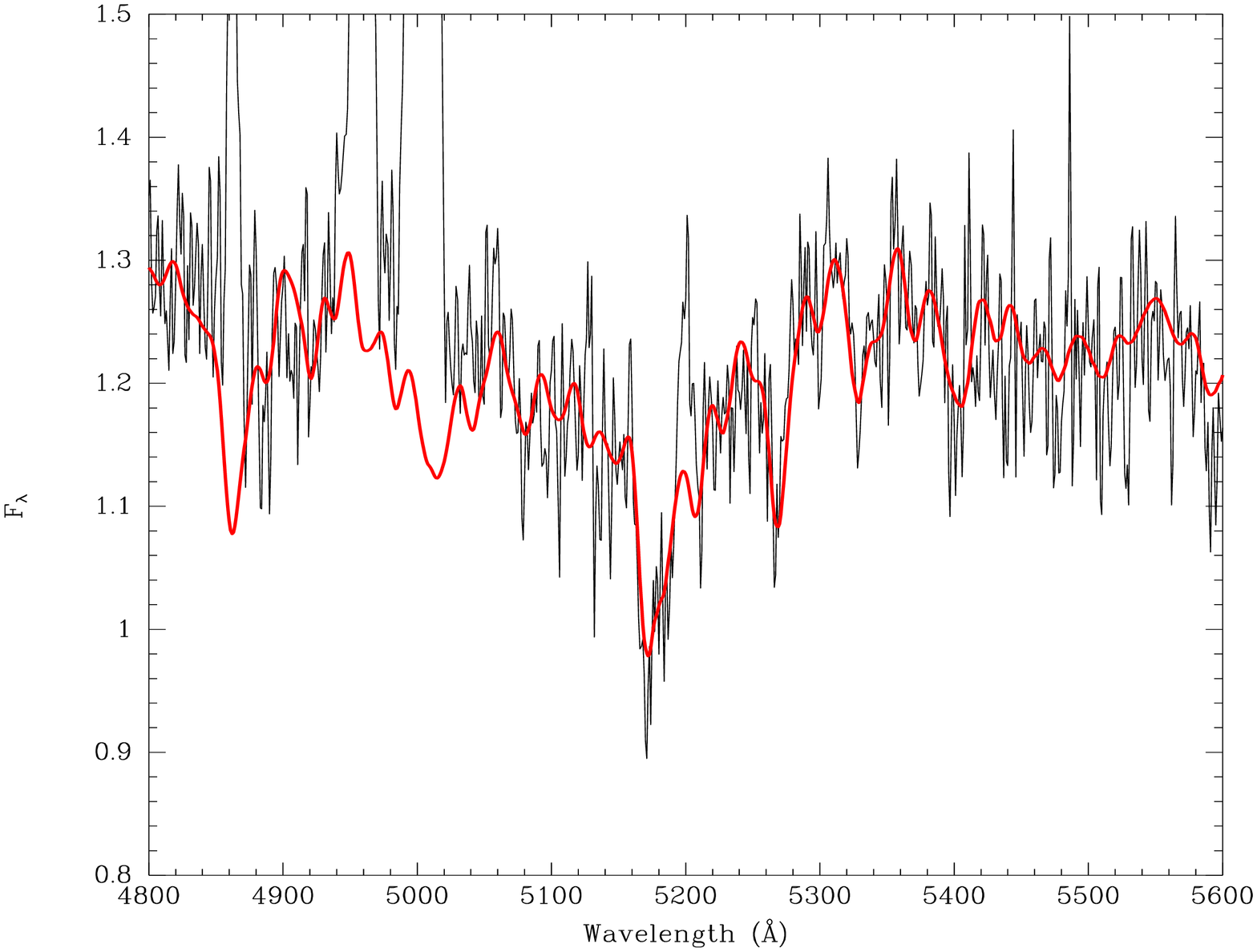}
\caption{Sample fit of the synthetic population model for SDSS
J150117.96+545518.2. Top: the observed spectra (top black curve,
shifted up for clarity), the synthetic spectra (middle red curve),
and the residual spectrum (bottom black curve). Bottom: the region
around Ca H+K $\lambda \lambda$ 3969, 3934 and G-band (left); the
region around Mg Ib$\lambda \lambda$ 5167, 5173, 5184 triplet
(right).}
\end{center}
\end{figure*}

\begin{figure*}
\begin{center}
\includegraphics[width=12cm,height=18cm,angle=-90]{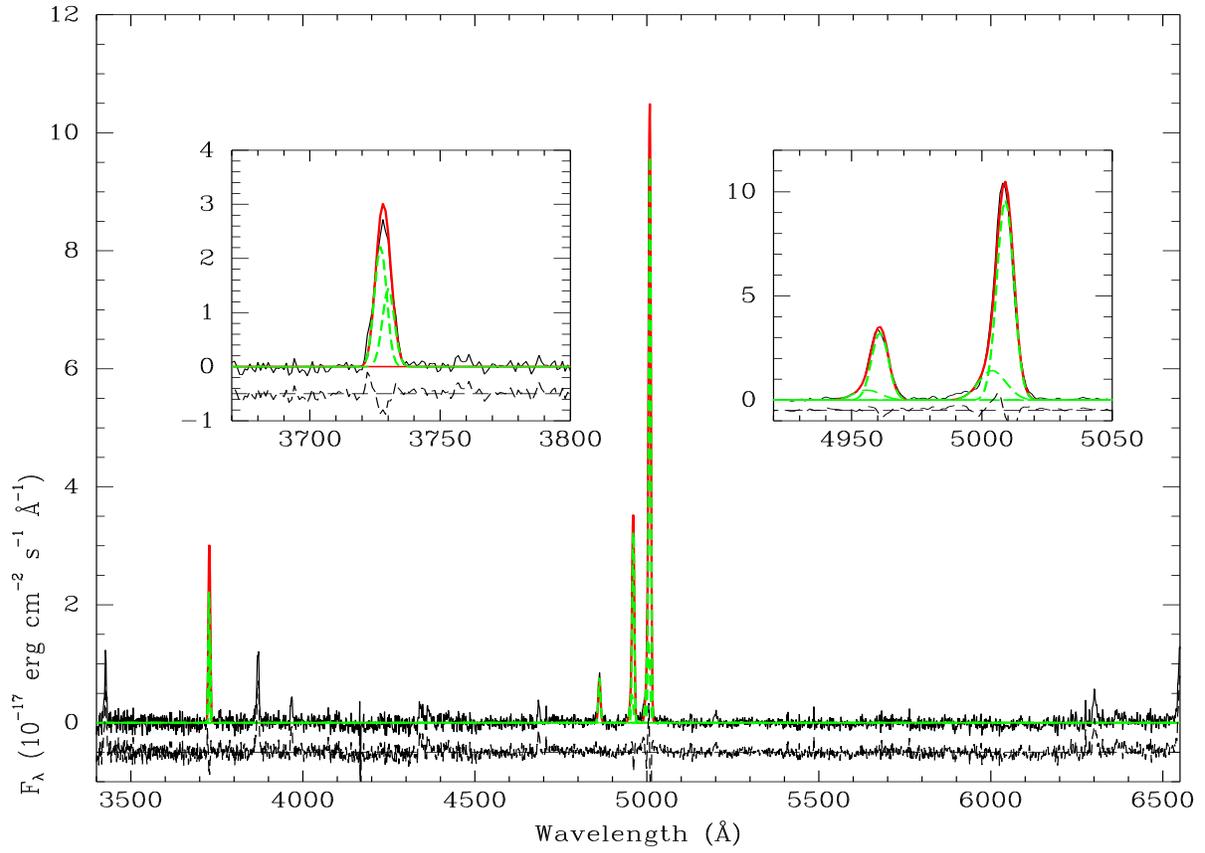}
\caption{Sample multi-component fitting of the [O
II]$\lambda\lambda$3726,3729 and [O III]$\lambda\lambda$4959, 5007
lines for SDSS J150117.96+545518.2: modeled composite profile
(thick solid red line), individual components (the dotted green
lines), the residual spectrum (lower panel).}
\end{center}
\end{figure*}

\begin{figure*}
\begin{center}
\includegraphics{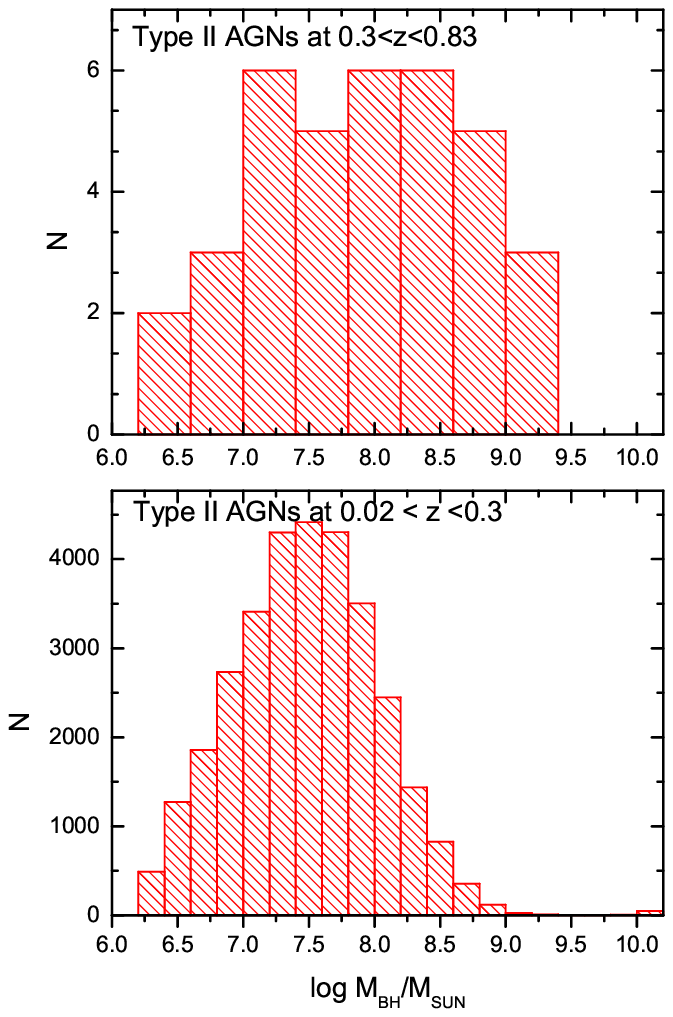}
\includegraphics{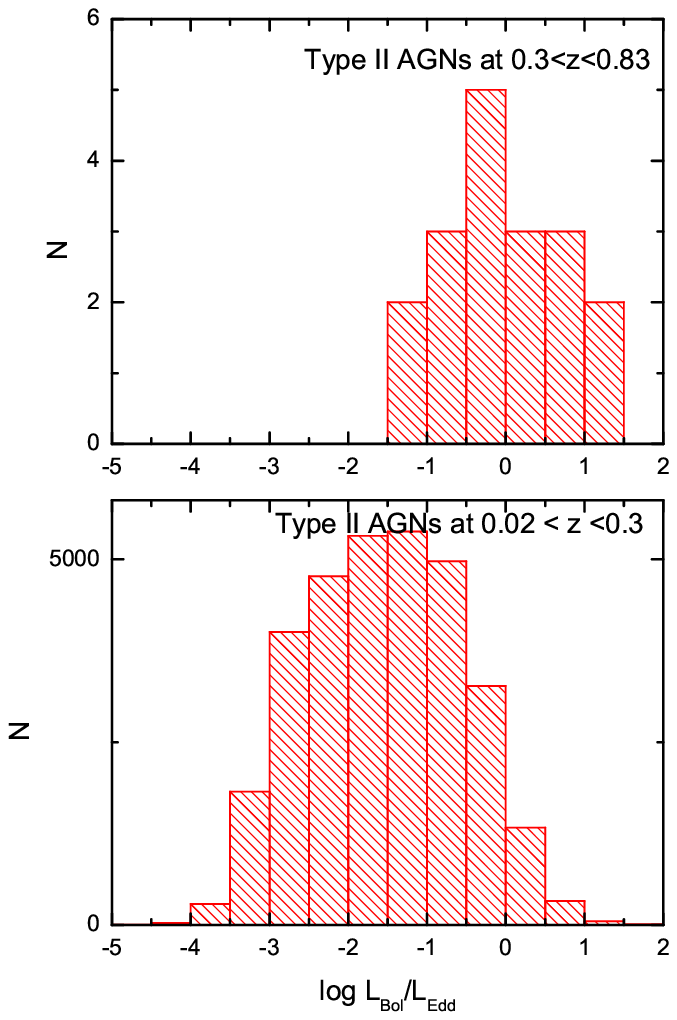}

\caption{The distribution of SMBHs masses and Eddington ratios for
our results (type II AGNs at $0.3 < z < 0.83$) and for Kaufmann et
al. (2003, type II AGNs at $0.02 < z < 0.3$)). }
\end{center}
\end{figure*}

\begin{figure*}
\begin{center}
\includegraphics[width=8cm,height=5cm]{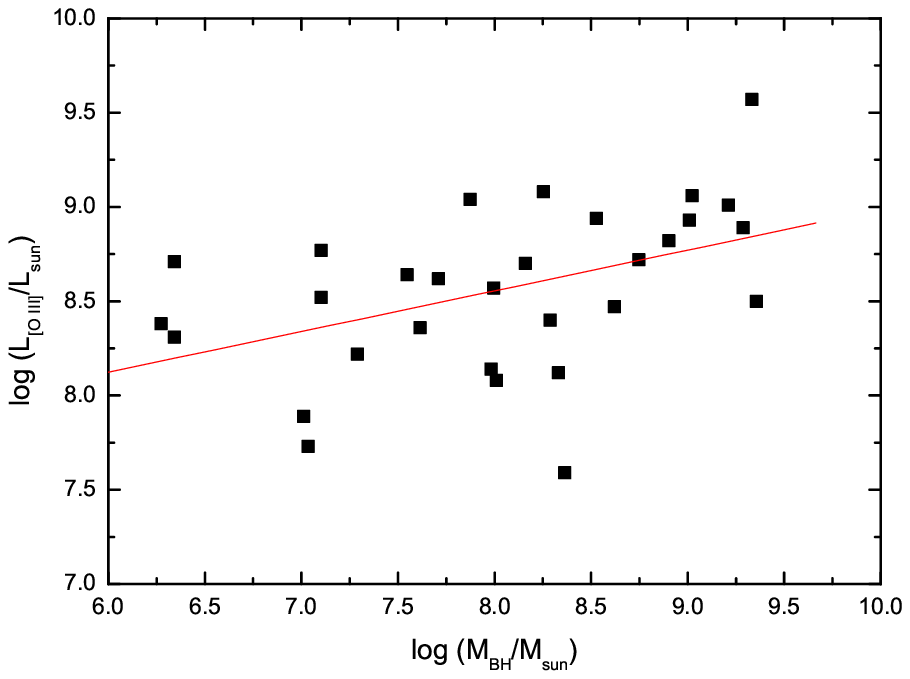}
\includegraphics[width=8cm,height=5cm]{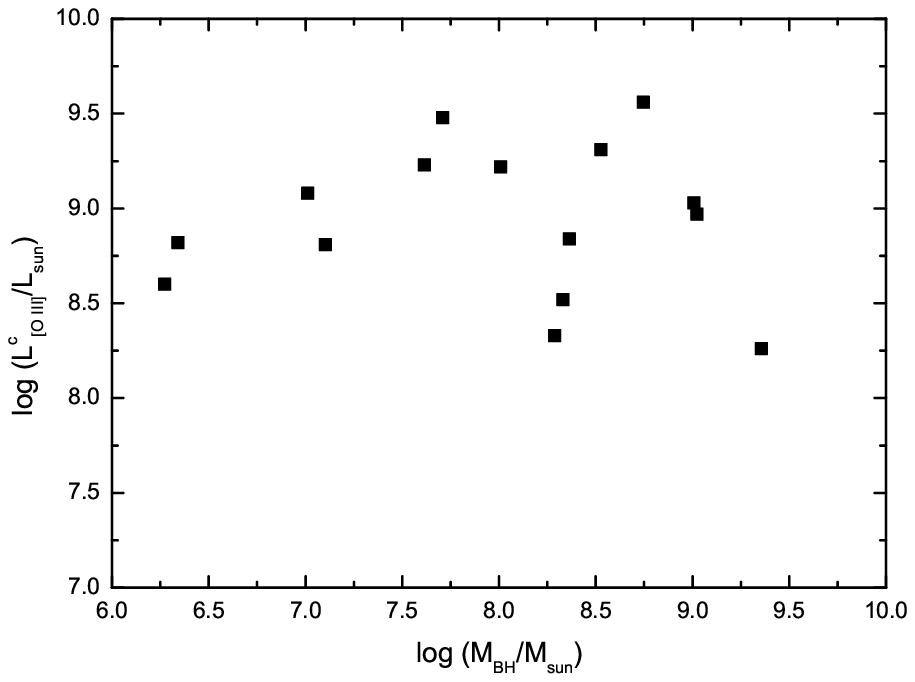}
\caption{The [O III] luminosity (left) and the corrected [O III]
luminosity (right) plotted against the SMBHs masses derived from
the stellar velocity dispersion $\sigma$. Solid line is the best
fit.}
\end{center}
\end{figure*}

\begin{figure*}
\begin{center}
\includegraphics[width=8cm,height=5cm]{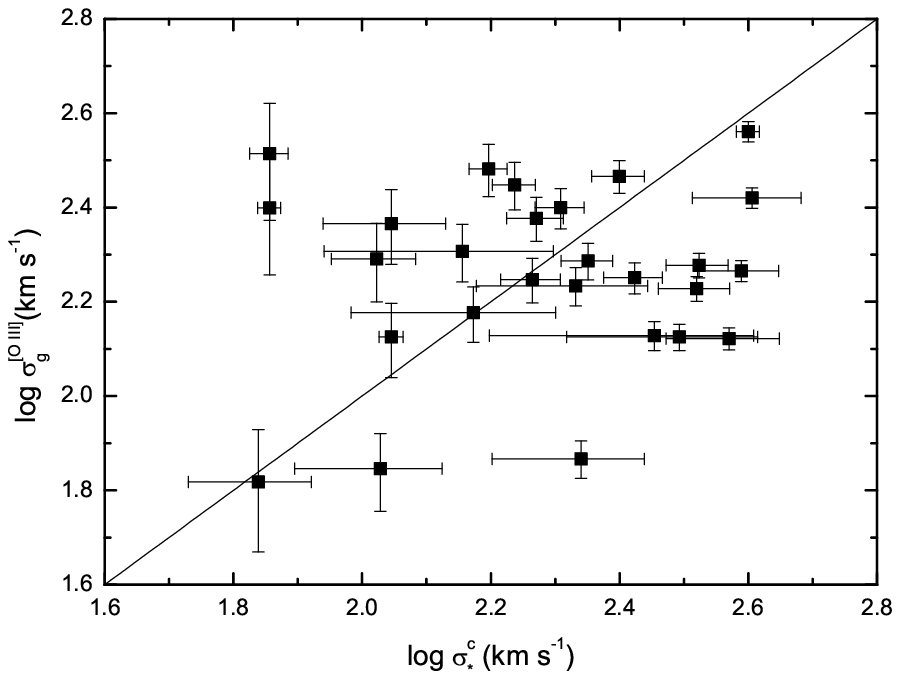}
\includegraphics[width=8cm,height=5cm]{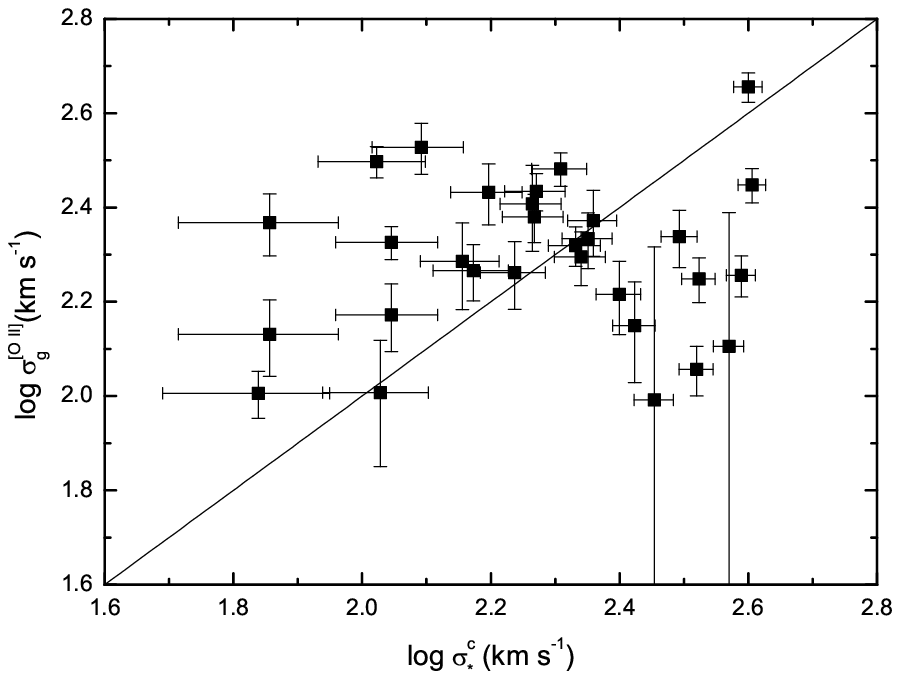}
\includegraphics[width=8cm,height=5cm]{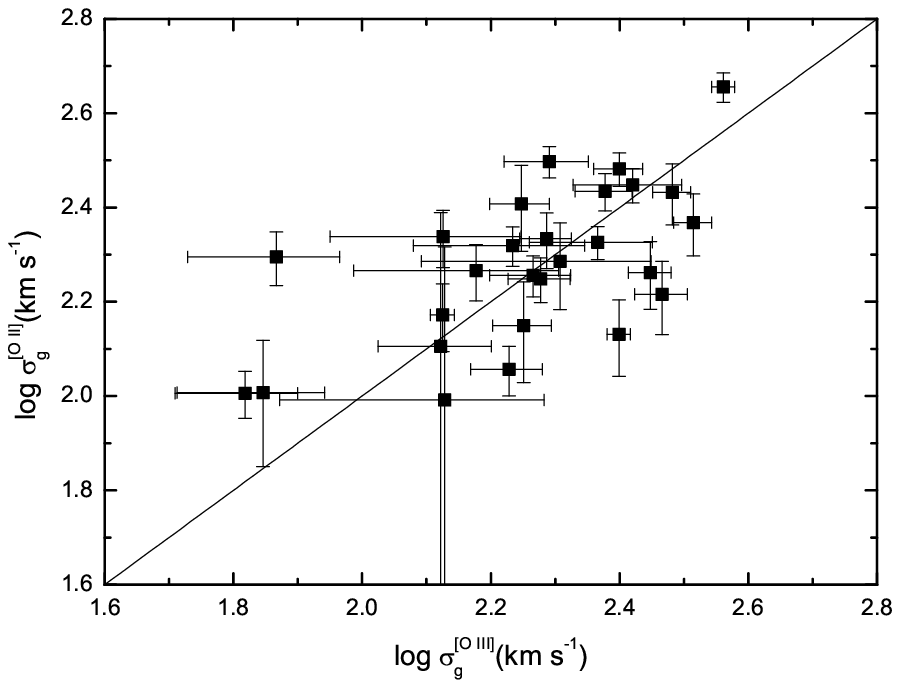}
\includegraphics[width=8cm,height=5cm]{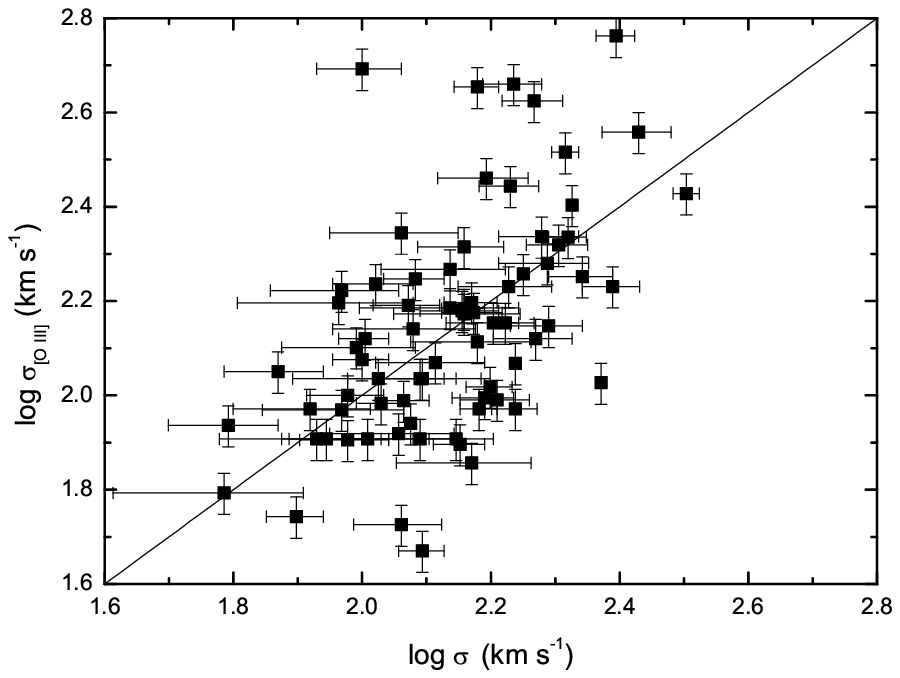}
\caption{Top left panel is $\sigma_{g}$ versus $\sigma_*^c$ for
the [O III] line; top right panel is $\sigma_{g}$ versus
$\sigma_*^c$ for the [O II] line; bottom left is $\sigma_{g}^{[O
III]}$ versus $\sigma_{g}^{[O II]}$; bottom right is $\sigma_{g}$
versus $\sigma_*$ for the the sample of Nelson \& Whittle (1996).
The uncertainty in $\sigma$ is typically 20 \kms. The solid line
denotes 1:1.}
\end{center}
\end{figure*}

\begin{figure*}
\begin{center}
\includegraphics[width=8cm,height=5.5cm]{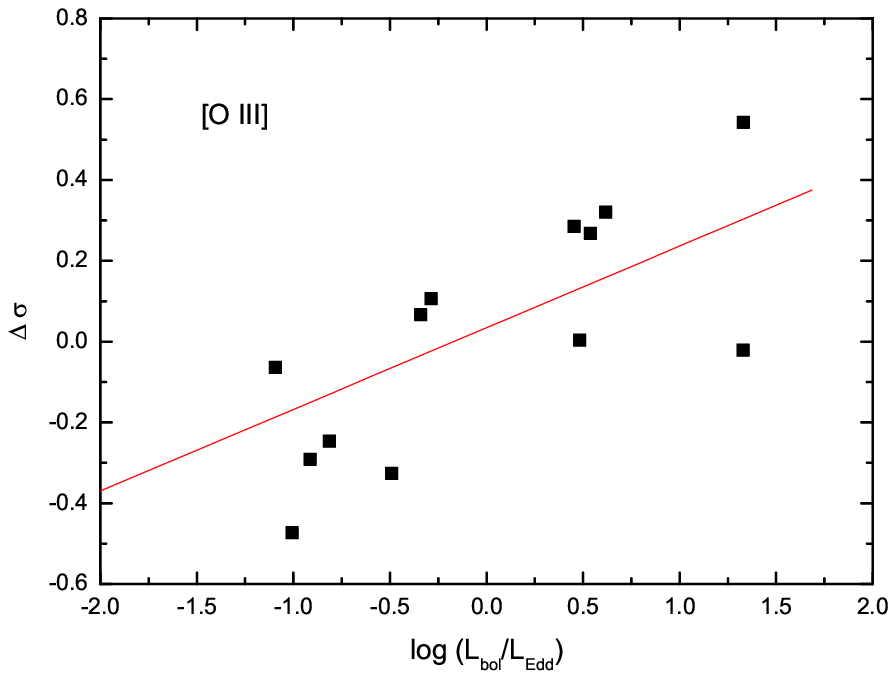}
\includegraphics[width=8cm,height=5.5cm]{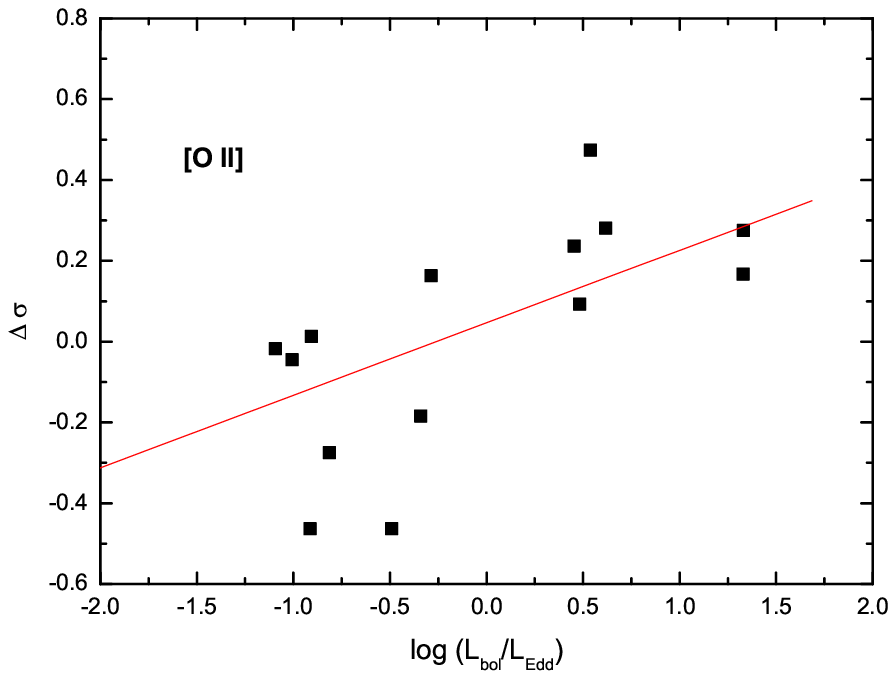}

\caption{$\Delta \sigma=log\sigma_{g}-log\sigma_*^c$ plotted
against the Eddington ratio, $L_{bol} /L_{Edd}$. Left panel is for
the [O III] line and right panel is for the [O II] line. Solid
lines show the best fits.}
\end{center}
\end{figure*}

\end{document}